\documentclass[aip,pop,reprint,groupedaddress]{revtex4-1}

\usepackage{amsmath}
\usepackage{lineno}
\usepackage{graphicx}

\begin{document}

\title{Plane strain optimization of conductor and structure grading in the inner leg of a Tokamak toroidal field coil}

\author{C.P.S. Swanson}
\email{cswanson@pppl.gov}
\affiliation{Princeton Plasma Physics Laboratory, Princeton University, Princeton, New Jersey 08543, USA}

\begin{abstract}

I present the results of the analytic plane strain optimization of the structure and conductor grading in the inner leg of a Tokamak toroidal field coil. The coil is assumed to be made of regions of soft conductor inside a stiff grid of conduit. Calculus of variations is used to determine the optimal profile of this structure. The optimal solution is found to be two-layered. An outer layer is bucked only, not wedged, and the structure fraction is graded so that all structure is at a uniform stress. An inner layer is an advanced bucking cylinder, similar to a Florida Bitter plate, whose radial stiffness is tuned so that its azimuthal stress is uniform. Such an advanced bucking cylinder would require advanced manufacturing to fabricate. These results should be seen an upper limits rather than achievable performance. Concepts that arise from this optimization, such as selectively softening layers of the bucking cylinder by drilling or cutting, may be of use to real designs in a less-optimized form. 

\end{abstract}

\maketitle

\section{Introduction}

Compared to the experiments of today, Tokamak-based fusion power plants will require a combination of higher magnetic fields\cite{greenwald_high-field_2018,greenwald_performance_2018,sorbom_arc_2015,coppi_optimal_2001,flanagan_overview_1986} and larger size,\cite{federici_demo_2018,corato_progress_2018} perhaps combined with lower aspect ratios.\cite{costley_towards_2019,menard_overview_2012,morris_mast_2012} Each of these contributes to an increased structural load on the inner leg of the Toroidal Field (TF) coil. 

For various reasons, one may wish to obtain the largest bore for a given magnetic field strength and outer radius, or the highest magnetic field strength for a given inner and outer radius, or the smallest outer radius for a given bore and magnetic field strength. To achieve this, it is common in Tokamak TF coil design to include radial gradation of conductor, and sometimes structure.\cite{bromberg_nested_1991,bromberg_aries-rs_1997,zhai_conceptual_2018,titus_fnsf_2021} However, to this author's knowledge, there has not been an effort to apply rigorous optimization to this problem. Such an optimal solution may be used as an upper bound on performance, and to inspire new engineering approaches to approach it.

To simplify the inner leg of the TF coil to the point that it is susceptible to analytic optimization, several assumptions are made. The model is detailed in Section \ref{sec:Model}; chiefly I make the assumptions of axial invariance (plane strain) and axisymmetry. I assume that the TF coil is made of a grid of stiff structural conduit, inside which soft conductor cable runs. Calculus of variations may be applied to this model, and the optimal radial profile of the structure-and-conductor grid dimensions may be determined.

In Section \ref{sec:Solution}, I find the solution to this optimization problem. The solution is split into two radial regions: an inner advanced bucking cylinder, and an outer bucked winding pack with a specific radial profile. 

The outer winding pack is bucked only (supported by radial pressure, not wedged azimuthally), reminiscent of early ARIES designs\cite{bromberg_magnet_1990,bromberg_aries-rs_1997} and the pre-scissor-key-failure design of ITER.\cite{huguet_iter_1993} The structure is graded such that at every radius its stress is \textit{just sufficient} to support the integrated Lorentz force density of all the conductor radially outward of it. 

The advanced bucking cylinder is described in Section \ref{sec:Bucking}. Where a solid bucking cylinder (cylindrical shell) is a feature of many TF coil system designs, the radial and azimuthal stress profiles produced by a solid shell feature a stress concentration at the inner radius.\cite{harvey_theory_1985,iit_kharagpur_design_2009,swanson_axisymmetric_2021} The advanced bucking cylinder described by the optimal solution has azimuthally-running slits which modify the radial stiffness so that the azimuthal stress is uniform, eliminating the stress concentration. This feature is similar to the cooling holes of a Florida Bitter plate magnet.\cite{gao_new_1996,bird_design_2004,nakagawa_detailed_1983} An advanced bucking cylinder may withstand \textit{twice} the external pressure ($\sqrt{2}\times$ higher magnetic field) of a traditional solid cylindrical shell before the inner bore $r_{In}$ must become zero.

Other efforts have been made to optimize the structure of TF coil systems. F. Moon identifies the smallest possible structural mass from the magnetic Virial theorem.\cite{moon_virial_1982} J. Schwartz uses a segregation of members in axial tension $\sigma_z > 0$ and transverse compression $\sigma_r<0,\sigma_\theta<0$. To this author's knowledge, the solution of a marginally-bucked winding pack and advanced bucking cylinder is novel.

Many engineering and practical considerations have been left out of this optimization. For example, the optimal profile requires machining an infinite number of infinitesimal holes into the structure. Clearly this is impossible, though it may be approached with advanced manufacturing. Likewise, out-of-plane forces on the TF coils exert a torque on the inner leg of most TF coil systems, which must be reacted by the structure. This requirement led to the failure of the earlier ITER TF coil design.\cite{huguet_iter_1993} This requirement is not considered in this analysis. Typical solutions to this problem include wedging the inner legs of the TF coil, or the use of flutes or keys between TF coils, or conformal ``caps" on the tops and bottom of the TF coil which limit access for maintenance.\cite{bromberg_magnet_1990,titus_structural_2003,titus_fnsf_2021}

The assumption that the conductor is soft compared to the stiff structure is not valid if high-temperature superconducting (HTS) tapes such as YBCO are used.\cite{fujishiro_database_2005} In this case, the optimal solution for the outer winding pack is invalid, though the advanced bucking cylinder may still be used. 

\section{The model}
\label{sec:Model}

\begin{figure}
\begin{center}
\includegraphics[width=0.6\linewidth]{ 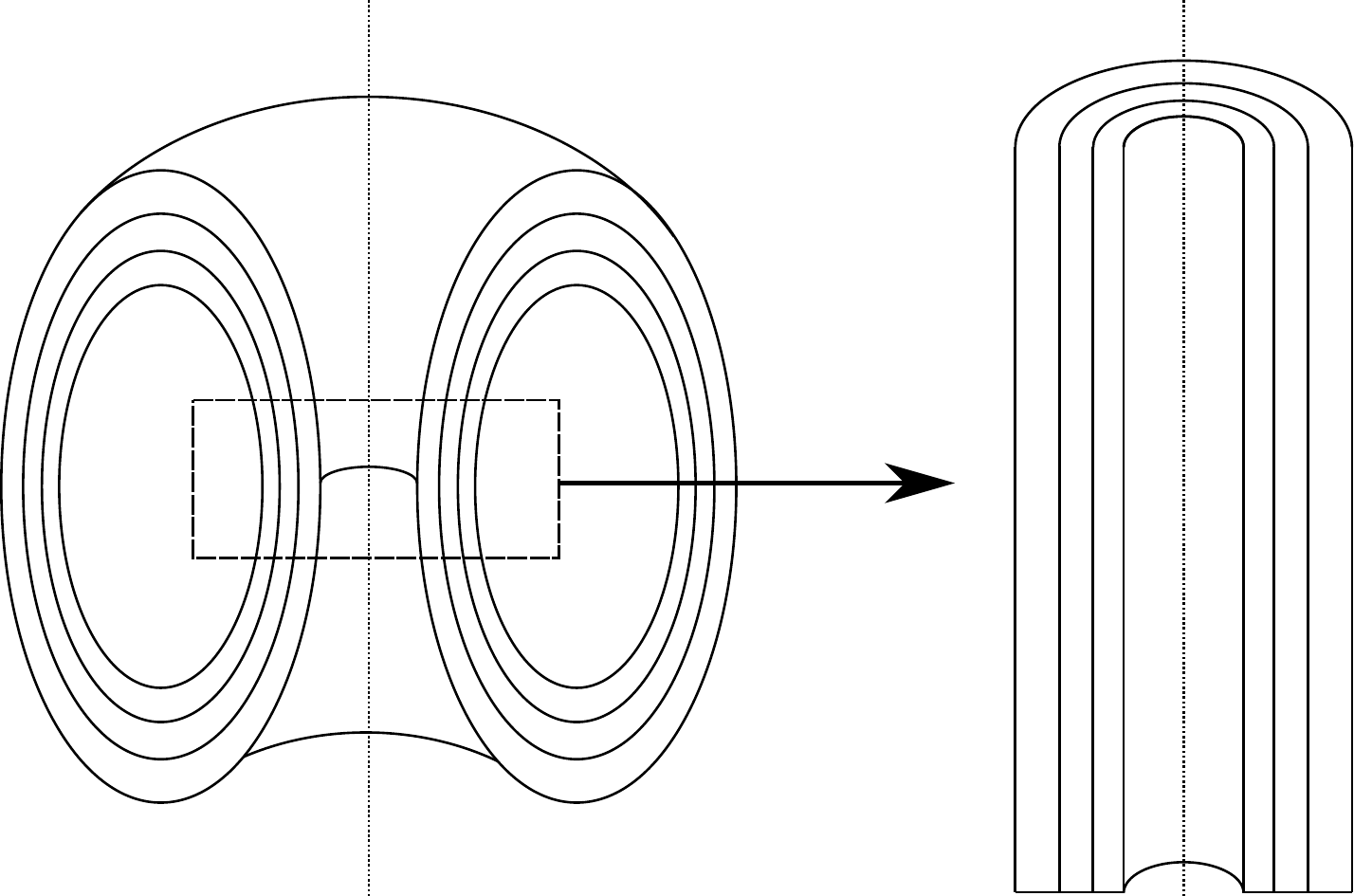}
\caption{The assumed geometry of the model: The inboard midplane of the TF coil system is assumed to map to an axisymmetric cylinder, long in $\hat{z}$}
\label{fig:Tok2Cyl}
\end{center}
\end{figure}

I assume that all structures are long in the axial $\hat{z}$ direction and that axial strain $\epsilon_z$ is uniform, see Figure \ref{fig:Tok2Cyl}. I assume that there are no torques or shears, sometimes called the ``generalized plane strain model" or more specifically the ``extended plane strain model."\cite{cheng_generalized_1995} I assume axisymmetry of the bulk structure, $\partial_\theta=0$. I assume that the Poisson's ratio of the structural material is zero, $\nu=0$. 

\begin{figure}
\begin{center}
\includegraphics[width=0.6\linewidth]{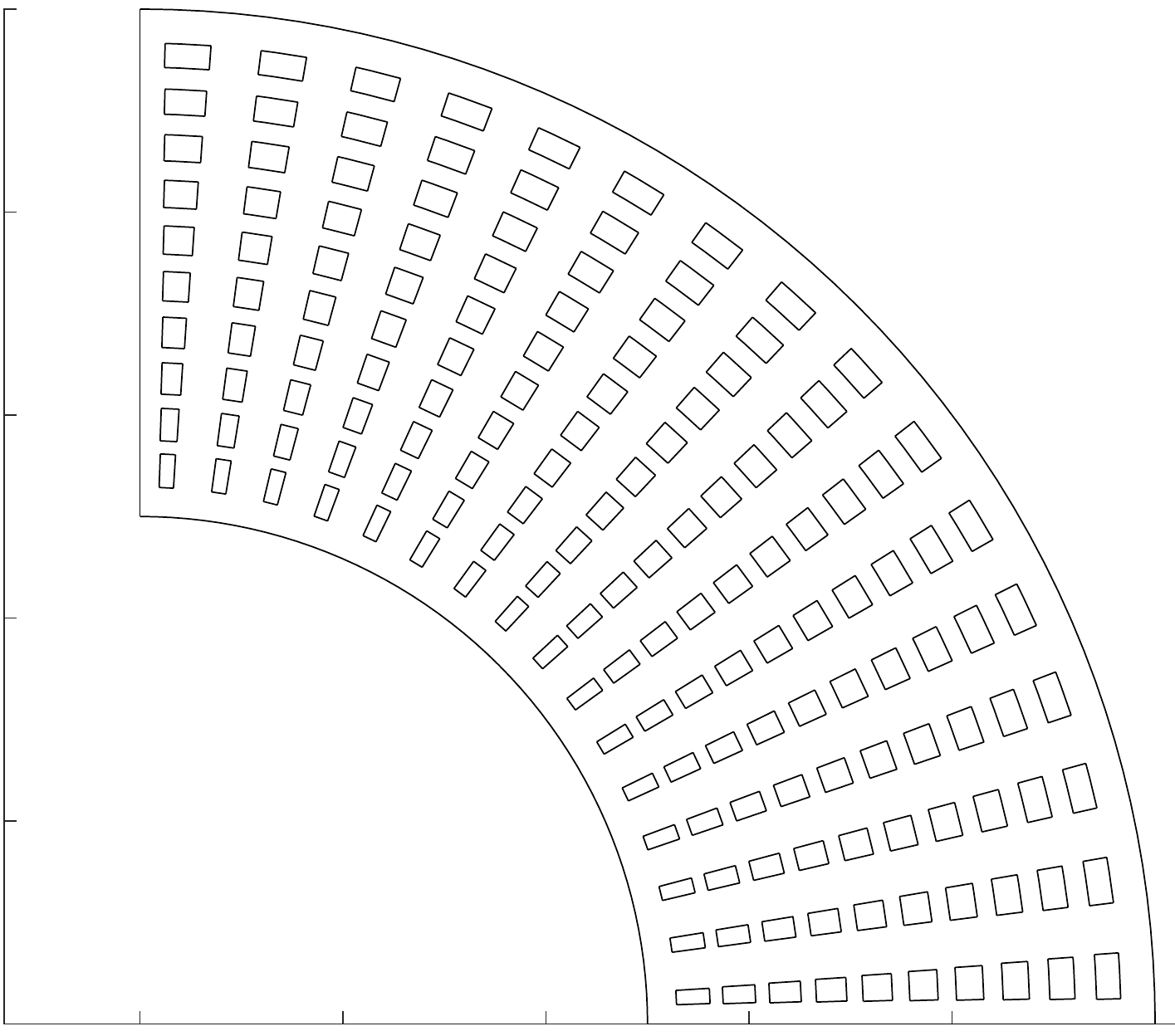}
\includegraphics[width=0.37\linewidth]{ 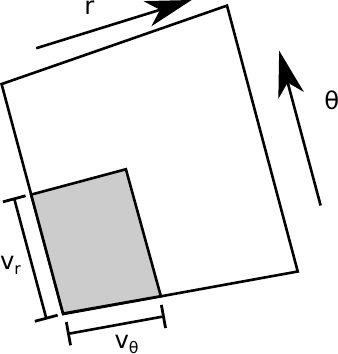}
\caption{Cross sections in $\hat{z}$ showing the micro-structure of the coil. Left: The coil is assumed to be made of a grid of stiff conduit, inside of which are axially-running conductors. Right: One grid cell of conductor and conduit. The conductor is colored gray. The linear void (conductor) fraction along the radial direction is $v_\theta$. The linear void fraction along the azimuthal direction is $v_r$. }
\label{fig:Conduit}
\end{center}
\end{figure}

I assume that the TF coil is composed of a grid of structural conduit, whose dimensions may change with radius. I assume that the Young's modulus (stiffness) of the structural material is much larger than that of the conductor, $E_{struct}\gg E_{cond}$. See Figure \ref{fig:Conduit}. The dimensions of the conductor region within the annular sector grid cell of conduit and conductor are characterized by linear void fractions $v_\theta,v_r$. Proceeding along a line that runs azimuthally through the conductor region, the fraction of this line which is void (conductor) is $v_r$. Proceeding along a line that runs radially, the fraction of this line which is void is $v_\theta$. $v_r(r),v_\theta(r)$ are functions of the radial coordinate. The reason the subscripts appear to be flipped is explained in Equations \ref{eq:EEffR} and \ref{eq:EEffT}.

The effective bulk Young's modulus of this micro-structured coil can be computed by the Voigt and Reuss property-smearing rules.\cite{voigt_ueber_1889,reuss_berechnung_1929,noauthor_derivation_nodate} The effective Young's modulus is anisotropic (orthotropic) and has the values:
\begin{equation}
E_{eff,r} = E_0 (1-v_r)
\label{eq:EEffR}
\end{equation}
\begin{equation}
E_{eff,\theta} = E_0(1-v_\theta)
\label{eq:EEffT}
\end{equation}
where $E_{eff}$ is the effective, bulk Young's modulus, and $E_0$ is the Young's modulus of the structural material. 

Force balance on a volumetric element in the axisymmetric plane strain model reduces to:
\begin{equation}
f = -jB = \frac{\sigma_{eff,\theta}}{r} - \frac{\partial_r (r\sigma_{eff,r})}{r}
\label{eq:fBalance}
\end{equation}
where $f$ is the Lorentz radial force density, $j$ is the axial current density, $B$ is the azimuthal magnetic field, $\sigma_{eff,r}$ is the effective (smeared) radial stress, and $\sigma_{eff,\theta}$ is the effective azimuthal stress. Effective quantities average over the micro-structure depicted in Figure \ref{fig:Conduit}.

The current density $j(r)$ can be found using the areal void (conductor) fraction $v_r v_\theta$:
\begin{equation}
j(r) = j_0 v_r v_\theta G(B)
\end{equation}
where $j_0$ is the critical current of the superconductor at zero magnetic field, and $G(B)$ is a critical current de-rating fraction such that the critical current at some magnetic field strength is $j_0 G(B)$. 

The first step in the optimization is to produce a family of non-unique solutions which are ``irreducible" in the sense that no material could be removed. To do this, we specify that the micro-structured conduit shown in Figure \ref{fig:Conduit} be uniformly at some maximum compressive stress:
\begin{equation}
\sigma_{micro,r} = -\sigma_{y,\perp}
\label{eq:SigMarginalR}
\end{equation}
\begin{equation}
\sigma_{micro,\theta} = -\sigma_{y,\perp}
\label{eq:SigMarginalT}
\end{equation}
that is, the micro-structure-resolved radial stress of the radially-running conduit $\sigma_{micro,r}$ and the micro-structure-resolved azimuthal stress of the azimuthally-running conduit $\sigma_{micro,\theta}$ are both in compression, and at some maximum yield stress $-\sigma_{y,\perp}$. 

This assumption does not produce a unique solution; rather we have down-selected from the space of all possible TF coil grid gradings to only those irreducible structures that use material in an irreducible way. The optimal solution is a member of this family, but this family includes off-optimal members.

The yield stress ($\sigma_{y,\perp}$) is given the transverse (``$\perp$") subscript in a concession to the fact that there is likely significant axial tension stress, $\sigma_z>0$. Applying the Tresca failure criterion, we find that the yield stress $\sigma_{y,\perp}$ must be de-rated by the axial stress:
\begin{equation}
\sigma_{y,\perp} = \sigma_{y,Tresca} - \sigma_{z}
\end{equation}
where $\sigma_{y,Tresca}$ is the yield stress in the Tresca criterion:
\begin{equation}
\sigma_{y,Tresca} = \textrm{max}(|\sigma_r-\sigma_\theta|,|\sigma_r-\sigma_z|,|\sigma_\theta-\sigma_z|)
\label{eq:Tresca}
\end{equation}

The axial tension stress $\sigma_z$ may be mitigated and even eliminated. This has been achieved by several methods in the literature and in practice, including combinations of compression rings at the top and bottom of the inner leg, slip-joints at the same location which do not transmit force, and even hydraulic pre-compression structures which compress the inner leg.\cite{titus_structural_2003,flanagan_overview_1986}

A common value of $\sigma_{y,Tresca}$ corresponding to common structural steels is 670 MPa.

Applying Equations \ref{eq:SigMarginalR} and \ref{eq:SigMarginalT} and the definitions of $v_r,v_\theta$ to the volume element force balance Equation \ref{eq:fBalance}, one obtains:
\begin{equation}
\partial_r v_r = \frac{v_\theta - v_r}{r} + \frac{j_0 v_r v_\theta G(B) B}{\sigma_{y,\perp}}
\label{eq:OptFamV}
\end{equation}

A differential equation for $B$ may be obtained from Ampere's Law:
\begin{equation}
\partial_r B = \frac{-B}{r} + \mu_0 j_0 v_r v_\theta G(B)
\label{eq:OptFamB}
\end{equation}

Equations \ref{eq:OptFamV} and \ref{eq:OptFamB} are a system of first-order Ordinary Differential Equations (ODEs) and may be solved between endpoints $r_{In}$ and $r_{Out}$ by any of the standard numerical methods. Solutions must also satisfy the constraints $0\le v_r \le 1,0 \le v_\theta \le 1$ and boundary conditions $v_r(r_{In})=1,v_r(r_{Out})=1$, this latter to ensure that there is no bulk radial stress at the edges of the coil.

Any choice of $v_r,v_\theta$ that satisfy Equations \ref{eq:OptFamV} and \ref{eq:OptFamB} may be said to belong to the irreducible family of TF coils, because the entire structure is uniformly at the maximum permissible compressive stress. There are an infinite number of such solutions. The next challenge is to find the unique optimal solution within this family, which may be solved via Calculus of Variations.

\section{The solution}
\label{sec:Solution}

I will now begin using dimensionless quantities and quantities most conveniently solved via Calculus of Variations. The quantities are:
\begin{equation}
x \equiv r/R_{mag}
\end{equation}
a dimensionless radius, where 
\begin{equation}
R_{mag} \equiv \sqrt{\frac{\sigma_{y,\perp}}{j_0^2 \mu_0}}
\end{equation}
a radial scale length of the problem.

$x_{Out}\ll 1$ corresponds to the case that the TF coil is more strongly limited by low current density than by structural requirements, and \textit{vice versa}.

We will maximize the magnetic field at the outer radius $B(r_{Out})$ for a fixed inner and outer radius $r_{In},r_{Out}$. This is equivalent to minimizing $r_{Out}$ at fixed $B(r_{Out}),r_{In}$ or maximizing $r_{In}$ at fixed $B(r_{Out}),r_{Out}$. The dimensionless, scaled magnetic field that we will maximize is:
\begin{equation}
L \equiv \frac{Bx}{\sqrt{\mu_0 \sigma_{y,\perp}}}
\end{equation}

The function to vary in order to optimize $L$ is the scaled linear azimuthal void fraction:
\begin{equation}
b \equiv xv_r
\end{equation}

With these quantities, we may re-write Equations \ref{eq:OptFamV} and \ref{eq:OptFamB} in this form:
\begin{equation}
\partial_x L = F(L,b,b^\prime,x) = \frac{b b^\prime G}{1+\frac{bLG}{x}}
\label{eq:F}
\end{equation}

Taking the first variation of all quantities around an assumed trajectory $Q + \delta Q$, we find that the solution for the variation is:
\begin{equation}
\begin{aligned}
\delta L(x_{Out}) = \int_{x_{In}}^{x_{Out}} dx^\prime \Big[\partial_bF - d_x \partial_{b^\prime} F + \partial_{b^\prime} F \partial_L F \Big] \delta b \\ e^{\int_{x^\prime}^{x_{Out}} dx^{\prime\prime} \partial_L F(x^{\prime\prime})}
\end{aligned}
\label{eq:Variation}
\end{equation}
The exponential is within the integrand. The derivation of Equation \ref{eq:Variation} is much the same as that of the Euler-Lagrange equation, except that a Green's function approach must be used because $F$ has a dependence on $L$. 

For the next step, to avoid tedious algebra, we will take $G(B) =1$, but this does not alter the conclusion.

Plugging $F$ from Equation \ref{eq:F} into Equation \ref{eq:Variation}, one obtains:

\begin{equation}
\begin{aligned}
\delta L(x_{Out}) = \int_{x_{In}}^{x_{Out}} dx^\prime \Big[ -\frac{Lb^2}{x^2(1+\frac{bL}{x})^2} \Big] \delta b \\ e^{\int_{x^\prime}^{x_{Out}} dx^{\prime\prime} \partial_L F(x^{\prime\prime})}
\end{aligned}
\label{eq:VariationSpec}
\end{equation}

Every factor multiplying $-\delta b$ in the integrand is positive-definite. Thus we conclude that, to maximize $\delta L(r_{Out})$, $\delta b$ must be \textit{as negative as possible} while satisfying the constraints. Equivalently, we may say that $v_r$ must be \textit{as negative as possible} while satisfying the constraints.

If we had included fully general $G(B)$ behavior into Equation \ref{eq:VariationSpec}, there would be another term multiplying $\delta b$ in the integrand proportional to $\partial_B G(B)$. Since this quantity is always negative for real superconductor critical surfaces, this consideration does not change the conclusion. 

The value of $v_\theta$ which allows $v_r$ to become the most negative the fastest while obeying its boundary conditions abruptly changes; it is $v_\theta=0$ during an inner layer and $v_\theta=1$ at an outer layer. Thus the solution is split into two radial regimes: 

A radially outward regime is characterized by $v_{\theta} = 1$. It can be numerically solved by iterating the coupled first-order ODEs Equations \ref{eq:OptFamV} and \ref{eq:OptFamB}, using $v_\theta=1$, until $B$ reaches zero at some radius. This solution corresponds to a winding pack. It is radially supported, in that there is no bulk azimuthal stress, $\sigma_{eff,\theta} = 0$. In TF coil design terminology, it is a ``bucked only," or ``non-wedged" winding pack.

A radially inward regime is characterized by $v_{\theta} = 0$. It can be analytically solved, having solution $v_r = r_{In}/r$. This layer carries no current; it serves only to react the radial body force applied to the outer layer. In TF coil design terminology, it is a ``bucking cylinder." However, it is not a traditional bucking cylinder, which is a solid cylindrical shell of structural material. Its radial Young's modulus has been tuned via azimuthal cuts into obeying the dependence $E_r \propto 1-r_{In}/r$. 

\begin{figure}
\begin{center}
\includegraphics[width=0.99\linewidth]{ 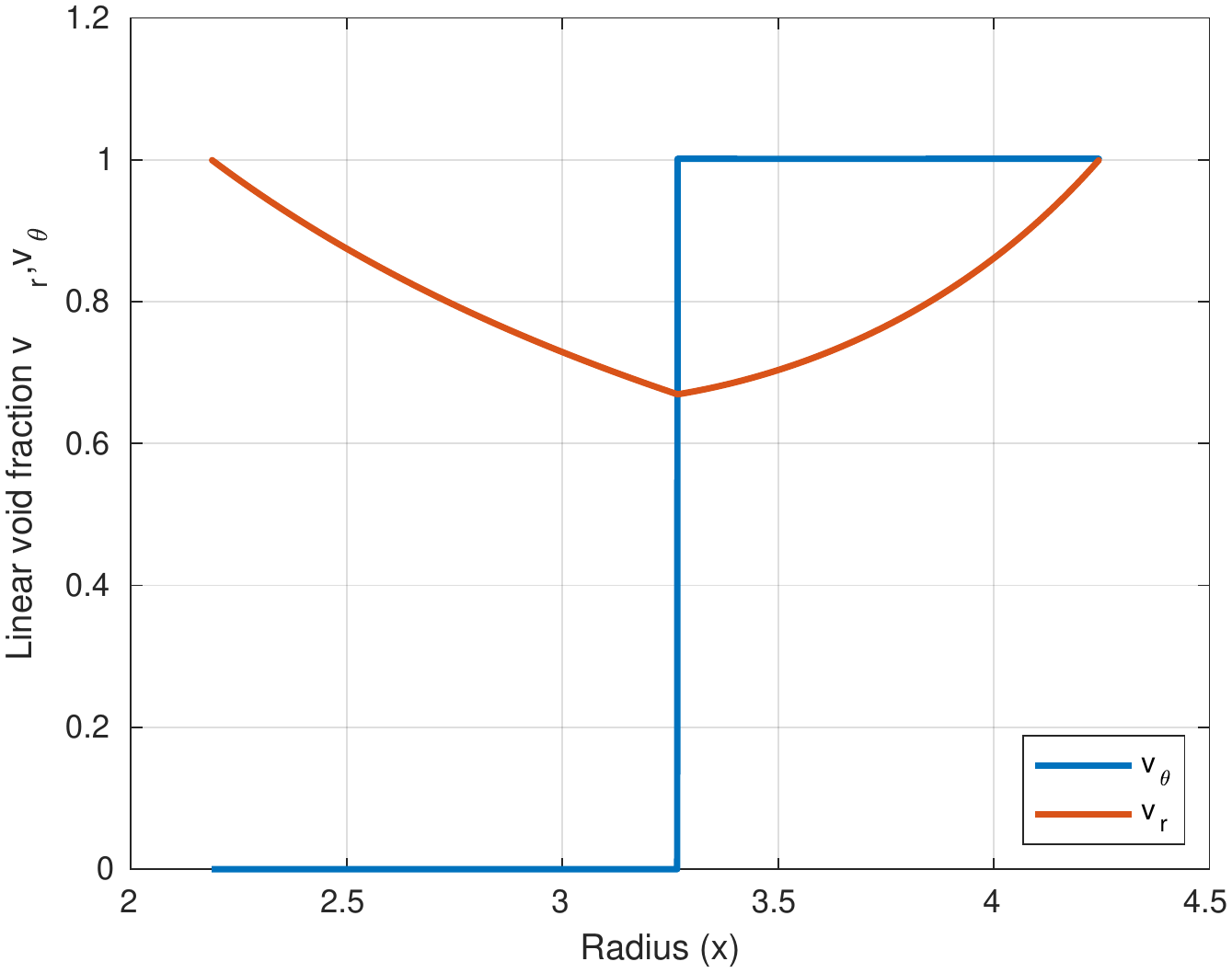}
\caption{An example optimal solution. The solution is split into an advanced bucking cylinder (inward of $\sim 3.25$, $v_\theta=0$) and a bucked-only winding pack with just enough structural to radially support conductors radially outward of it (outward of $\sim 3.25$, $v_\theta = 1$). This solution makes $v_r$ the \textit{most negative} it can be while satisfying constraints and boundary conditions.}
\label{fig:Example}
\end{center}
\end{figure}

An example solution can be seen in Figure \ref{fig:Example}. For this example solution, the values used were $x_{Out} = 4.24,L(x_{Out})=2.93$. The value of $x_{In}$ was numerically determined to be $2.19$. These dimensionless parameters correspond to a yield stress of $\sigma_{y,\perp}=670$ MPa, a current density of $j_0 = 140$ MA/m$^2$, an external magnetic field of $B(r_{Out})=20$ T, an outer radius of $r_{Out} =0.7$ m, and an inner radius of $0.36$ m. 

\begin{figure}
\begin{center}
\includegraphics[width=0.99\linewidth]{ 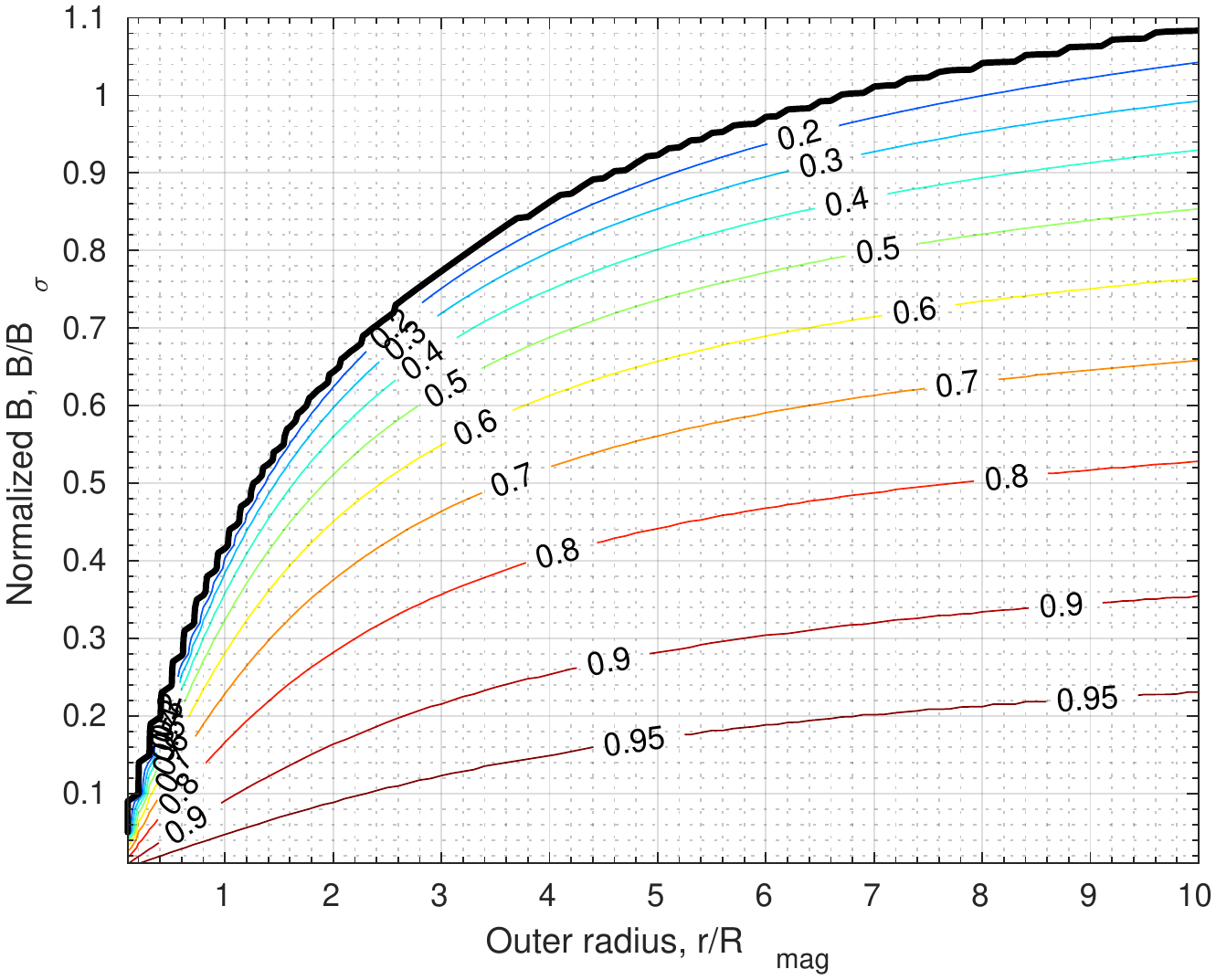}
\caption{Assuming no superconducting critical surface effects ($G(B)=1$), these are contours of the optimal (maximum) inner radius fraction $r_{In}/r_{Out}$, specifying an outer radius $x_{Out}= r_{Out}/R_{mag}$ and outer magnetic field $B(r_{Out})/\sqrt{\mu_0 \sigma_{y,\perp}}$ }
\label{fig:rIn}
\end{center}
\end{figure}

For reference, Figure \ref{fig:rIn} reports the values of $r_{In}/r_{Out}$ over a wide range of $x_{Out},B(x_{Out})/\sqrt{\mu_0 \sigma_{y,\perp}}$. These values may be used with the appropriate values of $j_0,\sigma_{y,\perp}$ to determine the optimal inner radius of a specific design. $G(B)=1$ was assumed here; for general $G(B)$, one must numerically solve Equations \ref{eq:OptFamV} and \ref{eq:OptFamB}, using $v_\theta=1$, until $B=0$. Inward of that, the solution is $v_r = r_{In}/r$, with $r_{In}$ computed via continuity of $v_r$. 

\section{The advanced bucking cylinder}
\label{sec:Bucking}

\begin{figure}
\begin{center}
\includegraphics[width=0.99\linewidth]{ 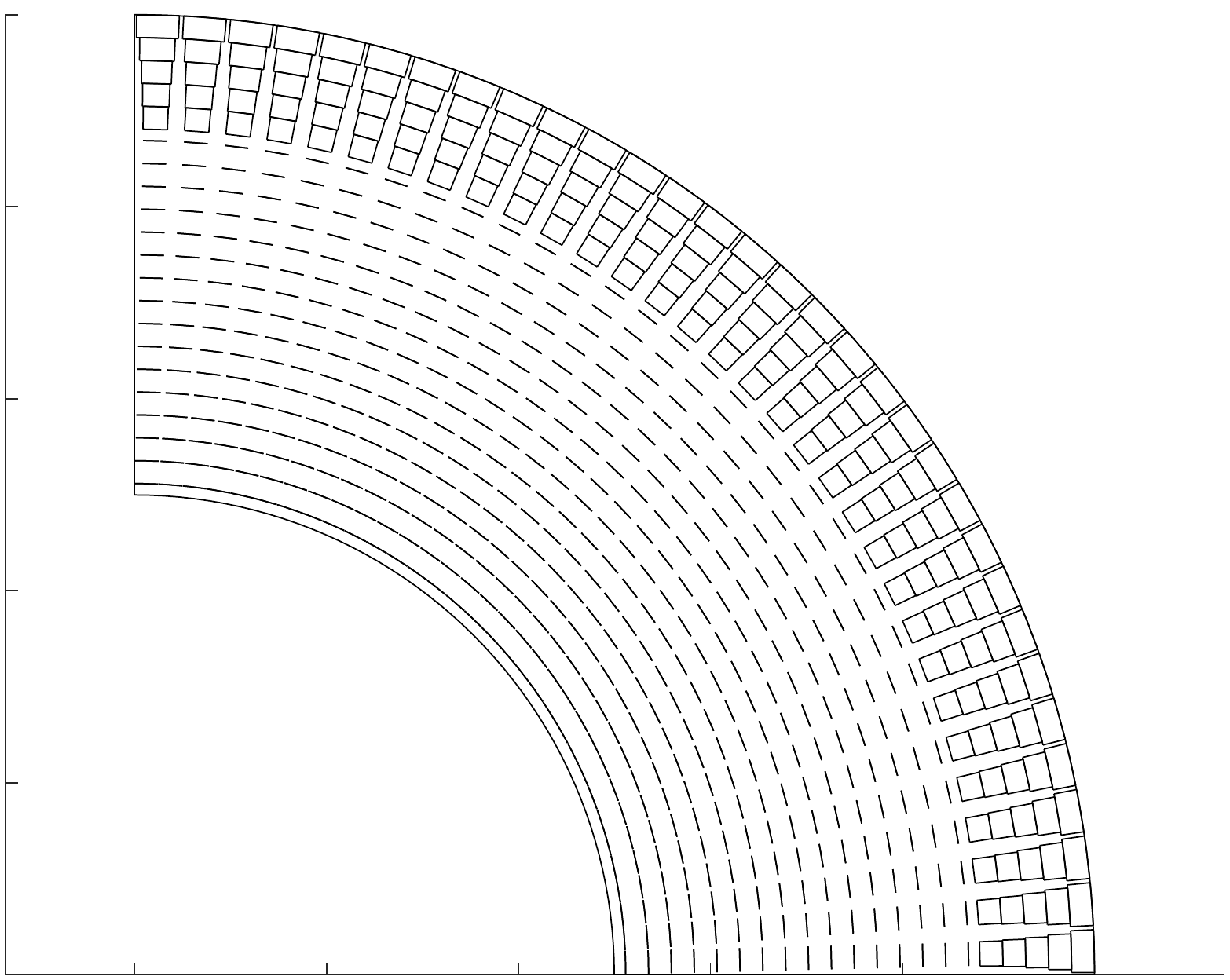}
\caption{A cross section in $\hat{z}$ of the optimal structure and conductor layout. There is an inner layer which carries no current; the cuts in the structure are only there to tune the radial Young's modulus, $E_r \propto 1-r_{In}/r$. This layer is the ``advanced bucking cylinder." There is an outer layer with a conductor fraction that thins as it goes radially outward; the thickness of these radial supports are just enough to support only those conductor layers radially outward of them. This layer is the winding pack.}
\label{fig:OptimalCuts}
\end{center}
\end{figure}

The inner layer of the optimal solution is an advanced bucking cylinder. It has azimuthal cuts that have been milled or drilled out of it, which tune its radial Young's modulus so that it has the dependence $E_r \propto 1-r_{In}/r$. The specific pattern of cuts can be seen in Figure \ref{fig:OptimalCuts}. The effect of these cuts is to decrease the radial stiffness, so that the azimuthal stress outward of them is increased and the azimuthal stress inward of them is decreased. In the limit of an infinite number of infinitesimal cuts, the azimuthal stress would be uniform, $\sigma_\theta = -\sigma_{y,\perp}$. 

These cuts to tune the radial Young's modulus are similar to the cooling holes in a Florida Bitter plate magnet.\cite{bird_design_2004} In a Florida Bitter plate, azimuthal cut-outs are drilled in the conductor in order to allow for cooling fluid to flow. Their placement and size are optimized to target equal temperature over the cross section of the coil.\cite{gao_new_1996,nakagawa_detailed_1983} These cooling holes have the benefit of making the azimuthal stress more uniform. The advanced bucking cylinder described in this work has similarities and differences; it does not require finite-area holes in order to admit cooling fluid, the bucking cylinder is long in $\hat{z}$ rather than a plate, the radial stiffness is fine-tuned to target uniform azimuthal compression, and the pressure that the bucking cylinder is withstanding is externally applied rather than volumetrically applied. 

The advanced bucking cylinder may be compared to a traditional bucking cylinder, which is a solid cylindrical shell of material. Stresses within the solid shell obey the thick cylinder Lam\'e pressure vessel profiles.\cite{harvey_theory_1985,iit_kharagpur_design_2009,swanson_axisymmetric_2021} The inner radius that a traditional solid shell bucking cylinder requires to withstand a pressure at some radius $P(r_P)$ is:
\begin{equation}
r_{In} = r_P \sqrt{1-2\frac{P}{\sigma_{y,\perp}}}
\end{equation}
where an advanced bucking cylinder may accomplish:
\begin{equation}
r_{In} = r_P (1-\frac{P}{\sigma_{y,\perp}})
\end{equation}

When $P\ll \sigma_{y,\perp}$, these cases are not appreciably different. However, when $P\sim \sigma_{y,\perp}$, the advanced bucking cylinder outperforms the traditional design by a wide margin. An advanced bucking cylinder may withstand \textit{twice} the external pressure of a traditional design before the bore $r_{In}$ goes to zero.

\section{``Virial efficiency"}
\label{sec:Virial}

The amount of structural material required can be related to a theoretical minimum provided by the magnetic Virial theorem.\cite{moon_virial_1982} I define the quantity ``Virial efficiency", $\epsilon_V$, as:
\begin{equation}
\epsilon_V \equiv \frac{\pi r_{Out}^2 B(r_{Out})^2/2\mu_0}{\sigma_{y,\perp} A_{Struct}}
\end{equation}
where $A_{Struct}$ is the cross-sectional area of the inner leg which is structural material, as opposed to bore or current-carrying conductor. 

For every member of the irreducible family of coils, $\epsilon_V \ge 0.5$. This can be seen because the pessimal member of the irreducible family, that of $v_r=1$, produces an analytic profile of $v_\theta$ whose $\epsilon_V$ is analytically evaluated at 0.5. 

I evaluated $\epsilon_V$ numerically for each optimal solution shown in Figure \ref{fig:rIn}. This analysis reveals that, for $B(r_{Out})/\sqrt{\mu_0 \sigma_{y,\perp}} \ll 1$, $\epsilon_V \rightarrow 0.5$. In words, if the magnetic field is not near the maximum possible magnetic field, the Virial efficiency of the optimal solution is not much better than that which can be obtained by other grading schemes.

However, as $B(r_{Out})$ increases and approaches $\sqrt{2 \mu_0 \sigma_{y,\perp}}$, the Virial efficiency increases and approaches $1$. That is, when the magnetic field approaches its maximum possible value, the Virial efficiency of the optimal solution approaches $100\%$ and is superior to other grading schemes.

\section{Conclusion}

To optimize the structure and conductor grading in the inner leg of a Tokamak Toroidal Field (TF) coil, I have carried out a Calculus of Variations analysis on a plane strain model. I assumed that the coil is composed of a stiff grid of structure, inside of which there is soft conductor. I found that the optimal grading of structure and conductor has two layers: An outer bucked-only (radially supported) layer, whose structure fraction is graded to be just sufficient so that each radial layer supports the conductor layers radially outwards of it, and an inner advanced bucking cylinder which is micro-structured to tune the radial Young's modulus so that the azimuthal stress is constant. 

I have produced a contour plot of the maximum possible inner radius (bore) for a range of outer radii and magnetic field strengths at the outer radius. These values may be approached but not exceeded by real TF coils.

The TF coil described herein is not manufacturable, as it requires an infinite number of infinitesimal micro-structuring operations. However, the result may be instructive as an upper limit on the performance of TF coils, and to inspire new engineering approaches. The approach of making the azimuthal stress profiles more uniform by cutting holes into the bucking cylinder may be useful to the field. 

\section{Acknowledgement}

I wish to acknowledge Y. Zhai and P.H. Titus of PPPL and S. Kahn of CCFE for very helpful discussions. I also wish to acknowledge PPPL engineers T.G. Brown, A.W. Brooks, and C. Rana, and PPPL physicist W. Guttenfelder.

This work was supported by the U.S. Department of Energy under contract number DE-AC02-09CH11466. The United States Government retains a non-exclusive, paid-up, irrevocable, world-wide license to publish or reproduce the published form of this manuscript, or allow others to do so, for United States Government purposes.

\section{References}

\bibliographystyle{plain}
\bibliography{ContOpt}

\end{document}